# NetRawALM: Network based Resource aware Application Layer Multicast for Multiparty Video Conference

T. Ruso[1] and C. Chellappan[2]

[1,2]Department of Computer Science and Engineering, Anna University, Chennai, India.

racingruso@gmail.com  drcc@annauniv.edu

## ABSTRACT

*IP Multicast is one of the most absolute method for large bandwidth Internet applications such as video conference, IPTV, E-Learning and Telemedicine etc., But due to security and management reason IP Multicast is not enabled in Internet backbone routers. To achieve these challenges, lot of Application Layer Multicast (ALM) has been proposed. All the existing protocols such as NICE, ZIGZAG and OMNI are trying to reduce average delay by forming a Multicast tree. But still that problem has not been addressed fully. We are proposing a new protocol called NetRawALM, which will address the average delay, Reliability between nodes, Scalability of conference, Heterogeneity and resilient data distribution for real time multimedia applications by constructing the Network based Resource aware Multicast tree algorithm. This is very dynamic and decentralised. The proposed architecture is a LAN aware; it is used to reduce Internet Traffic.*

## KEYWORDS

*IP Multicast, Application Layer Multicast, IPTV, Video Conference, Resource aware Multicast Tree, Scalability, Heterogeneity, Resilient.*

## 1. INTRODUCTION

An IP Multicast capable network allows one or more sources to efficiently send data to a group of recipients applications such as Video Conference, IPTV, E-Learning and Telemedicine etc.,. Here the sources transmit only one copy of the data and the appropriate network nodes efficiently make duplicate copies for each receiver. After a decade of research into the various issues of IP Multicasting such as routing, group management, address allocation, authorization and security, Quality of Service (QoS) and scalability, the widespread deployment of IP Multicast on the global inter-network has been dogged by technical, administrative and business related issues [3]. To avoid this problem we go for Multicast. Multicast has two types. One is IP Multicast and another one is Application Layer Multicast (Many to Many Unicast). IP Multicast is not being widely implemented so for. So that we approach Application Layer Multicast (ALM) it does not need any special support routing, forwarding, etc., from the underlying networks [2][20]. An overlay network is a computer network which is built on top of another network. Nodes in the overlay can be thought of as being connected by virtual or logical links, each of which corresponds to a path, perhaps through many physical links, in the underlying network.

The motivation behind approaching ALM, as opposed to the other proposed alternatives to IP Multicasting, is ALM's practical success and deploys ability on today's Internet, especially for home users, as demonstrated by file sharing applications such as Napster and Kazaa [2]. Every

ALM multicast tree rooted at the sender and participating nodes join the tree as interior and leaf nodes. An interior node is responsible for forwarding data from its parent node to its children through unicast [3]. There are three popular Application Layer Multicast protocols such as OMNI, ZIGZAG and NICE, approached in different way. They have tried to address the issues such as delay and jitter. Unlike network layer multicast where network packets are replicated at router inside the network, in ALM packets are replicated at end hosts. Logically, the end-hosts form an overlay network, and the goal of application layer multicast is to construct and maintain an efficient overlay for data transmission [1]. In order to provide the best quality of service we propose NetRawALM : Network based Resource Aware Application Layer Multicast Protocol. In this each peer in proportion to its available resources; low-delay, and fault tolerance, resource aware multicast trees are formed in the application layer in different networks. The tree formation procedure considers the number of clients, the available bandwidth, the available RAM, processor speed of the peers, hop distance to sort the various nodes that participate in a conference, resulting in a tree formation thus maximizing the quality of video and audio received at each node. On the other hand, a proactive approach takes into account the node departure before it happens. The basic idea is that each nonleaf node in the overlay multicast tree pre-computes a backup route. In Probabilistic Resilient Multicast (PRM) [18], each host chooses a constant number of other hosts at random and forwards data to each of them with a low probability. It enables each host to have a backup route. However, PRM generates extra data overhead. Another proactive approach was proposed by Yang et al [19], which we call Yang's approach in this paper. It calculates the *degree* each host has, and ensures backup route proactively whenever a node leaves or joins. *Degree* represents a outbound link. It is inevitable to consider the degree bound in overlay multicast, which can be easily observed in streaming applications. Each host limits the number of children on the tree it is willing to support [16][17].

We therefore propose a new proactive approach in order to avoid the degree limitation and generating heavy overheads [4][5][9]. By placing higher capability node in terms of available memory, processor speed etc., in the higher level of the tree. This method will produce the hierarchical structure tree. Using this tree, we can replace if any node failure in any level of the tree. Most of the video conference users are coming from the Local Area Network (LAN). The bandwidth of LAN is always same. So we are not considering this parameter to form the multicast tree. Furthermore, we implemented our proposal in software, and experimented with live video streaming over the actual network. The results of our implementation verify the effectiveness of our approach and convince us that our proposal achieved better streaming quality.

The rest of this paper is organized as follows. In section 2, we detail our related work. In section 3, we introduce Network based Resource aware Application Layer Multicast protocol. In section 4, we detail our Smart and Secure Environment Video Conference (SSEVC) application.  In section 5, we discuss some of the Performance Result and we conclude in Section 6.

## 2. RELATED WORK

There are lot of ALM protocols have been addressed so for, from those protocols, there are three popular protocols such as NICE, ZIGZAG and OMNI [10]. NICE protocol talked about two important factors such as stress, and stretch. Stress is the number of redundant packets that are send in one link. Stretch deals with per member the ratio of path length to the total path length. Layer formation in nice protocol, at first all the nodes are at layer 0. With some basic behaviour they form the cluster in layer 0, this time layer 0 has no of clusters. A centre node is selected as leader in each cluster and all the leader is moved to the layer 1. In this layer they form the cluster. From this cluster another leader is chosen and that leader goes to layer2. This is how it repeats until it form single node at top NetRawALM Protocol. The conditions are a node at some layer Lj must be present in the all the lower layer and a node which is not  in Lj should not be present in the Li where i>j. In a NICE hierarchy structure, members at that top have to maintain the state of the members which are at the lowest level and the members in the same group can only have some limited information about the other members in that group. While developing nice hierarchy the members which are closer to distance are considered to be same hierarchy. Nice hierarchy is created by assigning members to different layers [2].

If a node wants to join the tree, first it will query to the top layer one then the top node will give the information to the joining node about lower layer node. Then joining select one based on the rtt then sends to that node request. This is how they join the network when it comes to the lower layer that is layer0. There are two types of leaving one is Graceful: The leaving node informs other node of its departure and other one is Ungraceful: The leaving node "dies" suddenly without announcement. When nodes disappear from the tree, then we close all connections to the node and the method will inform to all other peers. Tree refinement is based on the following three methods: Cluster split, Cluster merge and Cluster leader transfer These refinements are done when the boundaries are violated. If it exceeds based on that merge or split or transfer can takes place. ZIGZAG protocol is similar to nice protocol while forming tree hierarchy but there are differences. First one is node on the same layer receives packet only from the higher node and a node on the same cluster receives from the same higher node. Second one is there is no connection or link between two nodes which are in the same cluster. Third one is the cluster nodes receives packet from the foreign node not from the parent node. This helps even if the parent node fails able to reconstruct easily. This is one of the advantages in ZIGZAG protocol. Here no scalability related things are considered. The node which wants to join to hierarchy should go from the parent node as similar to the NICE [2]. In OMNI protocol there is no leader and layered like structure as like in the previous two protocols. In this case service providers deploy multicast service nodes (MSN) that act as an application layer forwarding entities for set of client. OMNI helps in reducing the latencies to the entire client set. MSN are also assigned priority based on the number of clients that are attached to it. The main advantage is it minimizes the maximum latency [6][7].

## 3. NETRAWALM PROTOCOL

The main objective of this protocol is to increase number of participants (scalability), giving good Quality of service and achieve the heterogeneity of users. This protocol is consisting of three modules such as Resource Aware Multicast Tree Construction, Reachability Probability for Tree Refinement and Differentiated Service at IP Layer. A multicast tree is formed in the application layer with the participating nodes in the conference. The effective capacity of the participating nodes is considered to select a root of the tree. The children nodes communicate only with the root of the tree, which communicates with the other headers in different network, thus reducing the load on the server.

The Differentiated Service at the IP Layer includes modification of the Type of Service (TOS) field in the IPv4 or IPv6 header. The appropriate traffic class is set according to the precedence. This improves the quality of service of the streaming packets, thus reducing the delay and loss in transmission.

### 3.1. System Architecture

The main objective of the proposed system is the effective use of bandwidth. Based on the system properties of the client nodes in each network, the best node is chosen as the header node. A multicast tree is constructed to achieve this purpose [12][13][14].

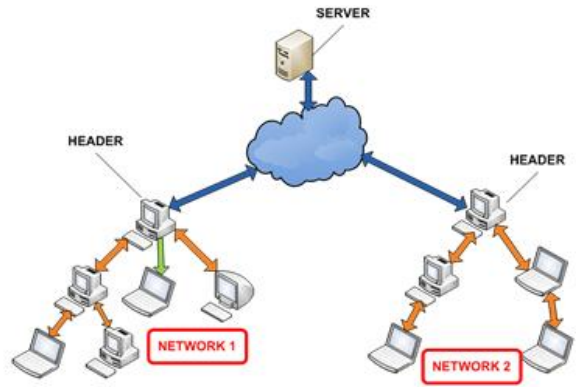

Figure 1: communication between different networks

In this above figure, there are two different networks are mentioned. In network1, among the available nodes a header (the best node) is chosen based on the system properties. The header is chosen in network2 in a similar manner. Based on the effective capacity of the nodes in the conference, the children nodes are added / configured to the root node, considering the application and network bandwidth. The server sends streaming packets only to the header nodes, which then transmit them to the children nodes. The double sided orange arrow indicates that the node is a participant (can participate in conference). The single arrow indicates that the node is a spectator (cannot participate in conference).

### 3.2. Resource Aware Multicast Tree Algorithm

The following are the pre-requisites for constructing the tree. The NETSTAT.exe in System32 folder can be used to find the Gateway and LocalIP addresses. Based on the gateway address, the nodes are identified to their networks. For each network, the best client is identified based on the following parameters. Available RAM – The system with maximum free RAM is selected and will transmit the streaming packets to the other nodes (children) at a comparatively higher speed. Hop Distance – This parameter is not required if all the nodes are at the same hop distance. CPU Speed – The header node should be faster so as to handle many children nodes. Processor info – This includes the number of processors, the processor architecture. The greater the number of processors the faster is the client.

The number of nodes in each level is determined based on the application and network bandwidth. The tree is constructed and the streaming packets are transmitted through the header only during a conference. The headers keep a track of the IP addresses and ports of the children nodes. Only the header nodes of each network communicate with each other and the server in a conference, thus effectively utilizing the bandwidth and minimizing the load on the server. The tree is formed only when the number of participants in the conference exceeds the desired size leading to packet loss. The following is the algorithm of the tree construction.

The following is the algorithm for configuring a node as the header. A header node maintains an arraylist of its children nodes. The children nodes are sorted using a priority queue and they are added to the tree in the order they occur in the priority queue. Initially the priority queue is sorted using a comparator based on cpu speed and available RAM.

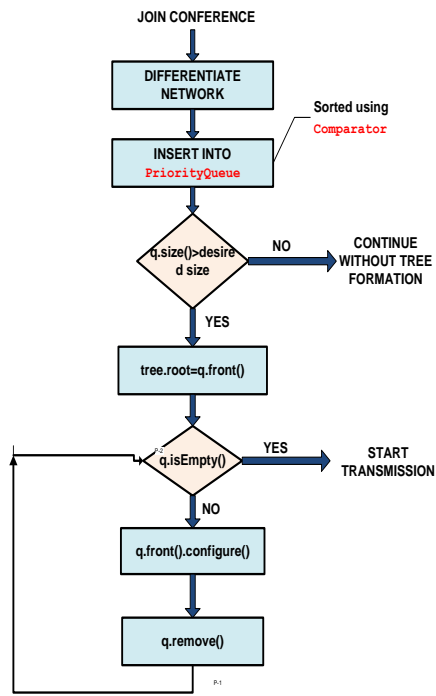

Figure 2: Algorithm for Tree formation

```
configureNode(PriorityQueue q){

    if(conditionsSatisfied){

    NodeDetails node=queue.peek();

    if(node.parent==null){

        this.children.add(node);

        node.parent=q.peek();

    }

    queue.remove();

    }

    else

        return;

    }

}
```

### 3.3. Differentiated Service

The Type Of Service(TOS) field in the IPv4 header is used to specify a preference for how the datagram would be handled as it made its way through an internet. A router maintains a TOS value for each route in its routing table. Routes learned through a protocol that does not support TOS are assigned a TOS of zero. Routers use the TOS to choose a destination for the packet.

The router locates in its routing table all available routes to the destination. If one or more of those routes have a TOS that exactly matches the TOS specified in the packet, the router chooses the route with the best metric. Otherwise, the router repeats the above step, except looking at routes who's TOS is zero. If no route was chosen above, the router drops the packet because the destination is unreachable. The router returns an ICMP Destination Unreachable error specifying the appropriate code: either Network Unreachable with Type of Service (code 11) or Host Unreachable with Type of Service. Thus the TOS field is modified accordingly to implement the differentiated services at the IP layer, thus reducing the delay and loss in streaming packets. The precedence used here is 32- Priority class.

| PRECEDENC | DECIMAL | DESCRIPTION |
| --- | --- | --- |
| 111 | 224 | Network Control |
| 110 | 192 | Internetwork Control |
| 101 | 160 | CRITIC/ECP |
| 100 | 128 | Flash Override |
| 011 | 96 | Flash |
| 010 | 64 | Immediate |
| 001 | 32 | Priority |
| 000 | 0 | Routine |

### 3.4. Member Join Process

When a new host joins the multicast group, it must be mapped to some cluster in layer. We illustrate the join procedure in Figure 2. Assume that host wants to join the multicast group. First, it contacts the Tree Head (TH) with its join query. The TH responds with the hosts that are present in the highest layer of the hierarchy. The joining host then contacts all members in the highest layer to identify the member closest to itself. In the example, the highest layer has just one member, which by default is the closest member to amongst layer members. Host informs of the three other members in its cluster then contacts each of these members with the join query to identify the closest member among them, and iteratively uses this procedure to find its cluster.

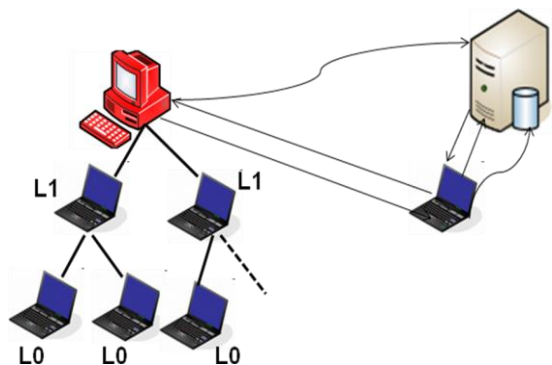

Figure 3: New member join

It is important to note that any host, which belongs to any layer, is the center of its cluster, and recursively, is an approximation of the center among all members in all clusters that are below this part of the layered hierarchy. Hence, querying each layer in succession from the top of the hierarchy to layer results in a progressive refinement by the joining host to find the most appropriate layer cluster to join that is close to the joining member.

### 3.5 Member Failure and Host Departure

When a host leaves the multicast group, it sends a *Remove* message to all clusters to which it is joined. This is a graceful-leave.

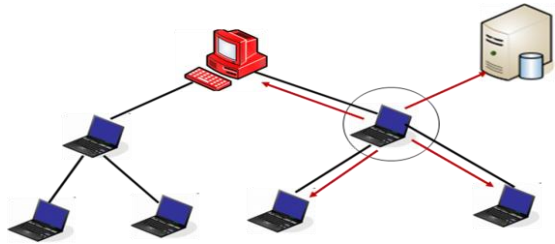

Figure 4: Member Departure

However, if it fails without being able to send out this message to all cluster peers, the algorithm will detects this departure through non-receipt of the periodic *HeartBeat* message from. A leader of a cluster, this triggers a new leader selection in the cluster. Each remaining member, of the cluster independently selects a new leader of the cluster, depending on who estimates to be the center among these members. Multiple leaders are re-conciled into a single leader of the cluster through exchange of *LeaderTransfer* message between the two candidate leaders, when the multiplicity is detected. It is possible for members to have an inconsistent view of the cluster membership, and for transient cycles to develop on the data path. These cycles are eliminated once the protocol restores the hierarchy invariants and reconciles the cluster view for all members.

## 4. SSEVC : SSE VIDEO CONFERENCE SYSTEM

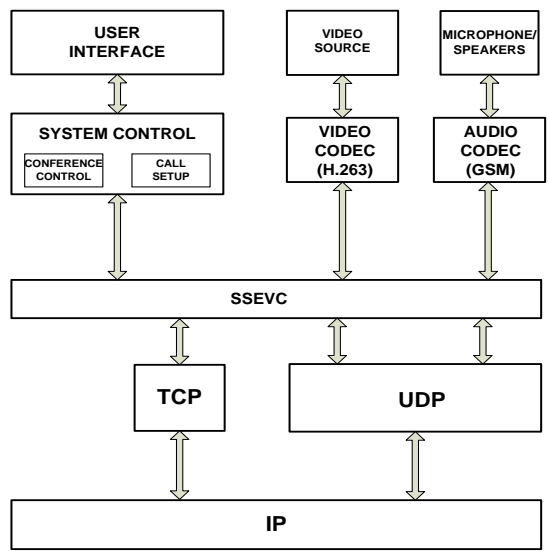

Figure 5: SSE Video Conference (SSEVC) System Model

The SSEVC protocol acts like an interface between Application layer and Network Layer. The User Interface includes the GUI provided that is, the Login screen, registration for a new user, options for entering the Server IP address and to select the mode of transmission: Peer-to-peer(P2P) or Server. The Conference Manager controls the Conference control and call Setup. The control messages like LOG_IN, CALL, JOIN_CONFERENCE_P, JOIN_CONFERENCE_S, LEAVE_CONFERENCE, END_CALL, CALL_ACCEPTED etc,. are transmitted through Control Packets through TCP layer. The audio/video data captured by the corresponding audio/video sources are encoded using codec's (Video codec: H.263 Audio Codec: GSM) and transmitted through Streaming Packets through UDP layer. The TCP/UDP packets are transmitted through IP layer which reach the destination.

## 4.1 Conference Management

Every user has to first register with the server by giving their user name and password. Once the user gives the username and password it is authenticated by the server. After successful authentication the server sends the list of on-going conference to the corresponding user, the user and chooses the conference of his preference then the server allows him to participate in that conference. After that the corresponding user's audio and video packets will be sending to all the participants in the conference. The user is allowed to create its own conference. Conference Management is discussed in the following sessions [11][15].

NetRawALM conference management protocol is a real time conference control protocol that is ready to use for everyday communications. It supports all types of Internet connections, including LAN, broadband, and even dial-up. It can be integrated with all kinds of instant messaging services, and a version for MSN messenger has been developed. The full mesh conferencing structure is first introduced in [2], where Lennox et al. also point out that the full mesh conferencing architecture is not suitable for bandwidth-limited end systems, such as wireless devices and users with 56 kbps modems. To break this limitation, in our system, we entirely separate the transmission module from the media stream engine and define a whole set of APIs that are open for both Unicast and application-level multicast (ALM).

When there are multiple data receivers, multicast allows data replication to be performed outside of the data source. Application-level multicast is different from traditional IP multicast in that data replication is conducted at end systems instead of multicast-enabled routers. With a proper ALM algorithm, we are able to alleviate the scalability problem of full mesh conferencing architecture. Our protocol is designed based on the full mesh architecture, where conference members are united by a fully connected communication mesh. And all the members are equivalent in terms of position in topology or rights in the conference. Different from, our protocol is so concise that it uses only four communication messages [2]:

♦ **JOIN_CONFERENCE as a Participant**:

A peer can join a conference as a Participant only if it is in the list of allowed participants specified by the conference Host. A peer intending to join a conference as a Participant generates a (private key, public key) pair and sends a JOIN_CONFERENCE message containing the conference name and its public key to the Server. The Server checks the list of allowed participants for that conference and if the peer is allowed then it sends the public key of the peer to other Participants and Spectators and also sends the public key of other Participants to the peer. Now the peer can subscribe to multicast AV data (encrypted) of other Participants and decrypt it before rendering using corresponding public keys. Also other Participants and Spectators can subscribe to the multicast AV data (encrypted) of the new Participant and decrypt it before rendering using the public key of the new Participant.

♦ **JOIN_CONFERENCE as a Spectator**

A peer can join a conference as a Spectator only if it is in the list of allowed spectators specified by the conference Host. A peer intending to join a conference as a Spectator sends a JOIN_CONFERENCE message containing just the conference name to the Server. The Server checks the list of allowed spectators for that conference and if the peer is allowed then it sends the public key of all Participants to the Spectator. Now the new Spectator can subscribe to the entire Participants' multicast AV data (encrypted) and decrypt it before rendering using the corresponding public key.

♦ LEAVE_**CONFERENCE as a Participant**

The Participant sends a bye message LEAVE_CONFERENCE containing the conference name to the server. The Server sends messages to all the Spectators and other Participants saying this particular Participant has left the conference . The Spectators and other Participants unsubscribe to the ex-Participant's multicast data.

♦ **LEAVE_CONFERENCE as a Spectator**

The Spectator can silently leave the conference unless the server wants to maintain a log of conference activities, in which case the Spectator sends a bye message LEAVE_CONFERENCE containing the conference name to the Server and the Server appropriately logs the incident.

### 4.3 Peer Life Cycle

Every peer has to get authenticated by the Authentication Server using its username and password. The server stores the IP of the peer's machine which can be used by other peers. On successful login, the system starts audio/video capture and renders only video locally. It continuously listens over a port for UDP update messages from the server. For e.g. list of online peers, list of on-going conferences. The peer can interact with other peers or join a conference [7]. On logout, peer should send a "BYE" message to the server and exit gracefully.

### 5. EXPERIMENTAL RESULT

Let P be the maximum no of nodes that can participate in the conference without tree formation. Let us assume that N nodes are participating in the network. If N<P means then tree formation is not required. If N>P means then tree is constructed. Height of the tree h is determined by the network and application bandwidth. The strength of a level (number of nodes in the level) is given by $2^{(h-1)}$. P is calculated based on the free network bandwidth and application bandwidth. P=free network bandwidth / application bandwidth For example, Let V1, V2, V3, V4 be the nodes participating in the conference. Let V1 has 1MB available RAM and 512MHz processor speed. V2 has 3GB available RAM and 2.00GHz processor speed. V3 has 4GB available RAM and 2.37GHz processor speed. V4 has 1GB available RAM and 1.2GHz processor speed. Let us assume that network bandwidth capacity is 2Mbps. free network bandwidth available is say 512Kbps. suppose our application bandwidth is 250 kbps. Then p based on above formula is 2.But here 5 nodes are there (N>P). So tree should be formed. These nodes are sorted based on their effective capacity (by the means of parameters already discussed) by a priority queue. The node which is first in the queue is chosen as the header.

Hence the head of the queue is V3, and the following nodes in the queue are in the order, V2, V4, and V1. V3 will be the root of the tree. The children nodes of V3 will be V2, V4 and V1.

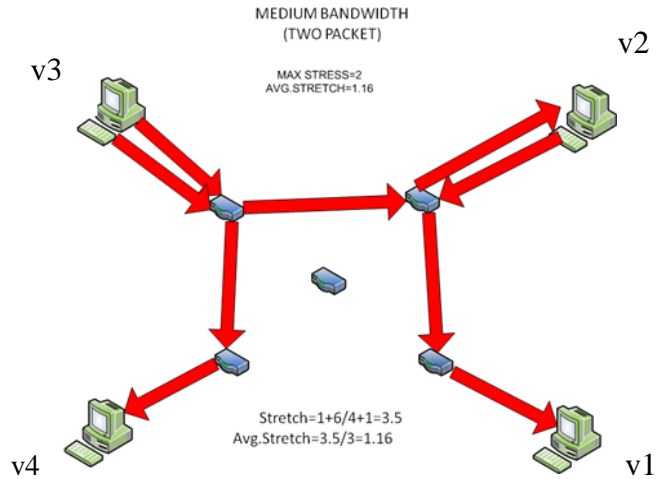

Figure 6: Ring multicast

Suppose network bandwidth is have some high capacity such that two packets can send at a time. In this case stress on the topology is two. The stretch experienced by different members are 1,6/4=1.5 and 1. Thus in this case where the maximum stress is 2 while the average stretch is reduced to 1.16

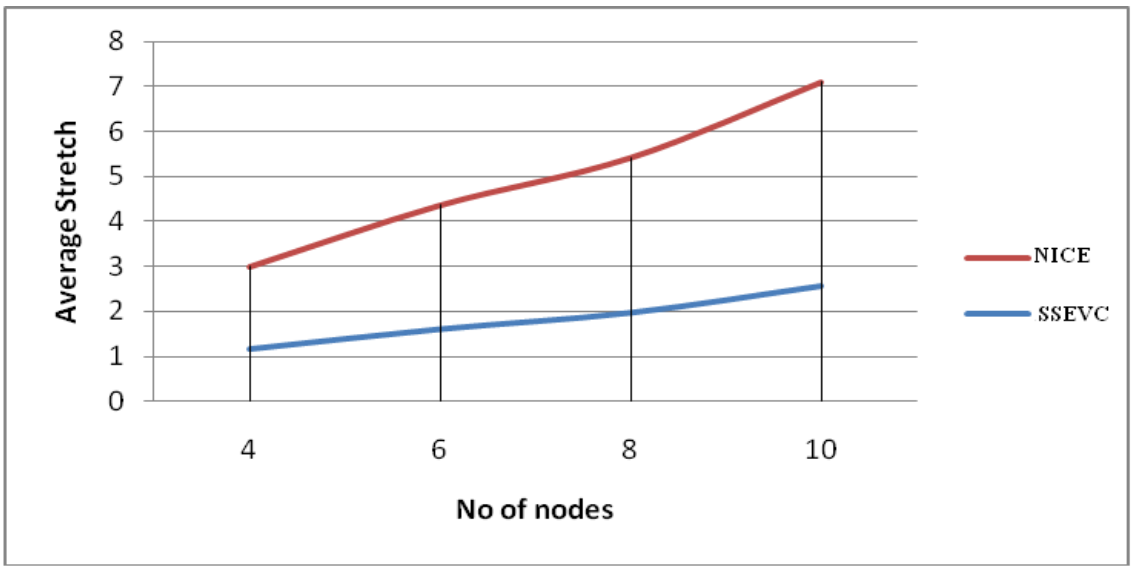

Figure 7: Comparison of NICE and SSEVC

The above graph is plotted between the AVG STRETCH on y-axis and NO.OF.NODES on x-axis of NICE AND SSE protocol. In NICE protocol only one packet will be send from the source node. Packet will reach the one node. Then from that node it will be send to another node. These how NICE protocol works. Here delay which is stretch calculated ratio of length of path the packet took to reach that node from the source node to the shortest length it can reach the node form the source node. Average of all the stretch is taken and it gives the average stretch which is plotted for different cases. Average stretch has taken for 4 nodes, 6 nodes and so on and they are plotted in graph. Even if the bandwidth is high enough to send two packets only one packet is send. In SSE we are sending two packet if bandwidth is high the average stretch gets decreased which is shown in the graph. Average stretch decreased is we send two packets in all cases for 4 nodes, 6nodes and so on. Thus the average delay gets decreased when we compared to NICE. The average stretch between SSE and NICE are shown in the above graph for different number of nodes. This protocol was tested using four different network bandwidth from various institutes such as SSE lab Anna University, Network System Laboratory IIT Madras, SSE lab Pondicherry University and SSE lab NIT Tiruchy.

## 6. CONCLUSIONS

In this paper we have presented NetRawALM protocol, which is significantly improving average end-to-end delay and it, support to increase the number of participants in the conference. Different Users can participate in the conference with different Heterogeneous resources such as bandwidth, devices and computing power. Reachability Probability is used to achieve Reliability between the hosts. Compare to the Existing protocol NICE, our protocol is reduced the average stretch between hosts. Our SSEVC application is very user friendly and it has the facility to control the QoS. We have created the overlay network test bed and we have done Differentiated Data Distribution over the test bed. In future we are planning to introduce Middle ware in each clusters, it monitor the QoS parameter of each cluster. This method will be used to improve the performance of the overlay test bed.

## ACKNOWLEDGEMENTS

First of all, we must acknowledge Collaborative Directed Basic Research in Smart and Secure Environment (CDBR-SSE), National Technical Research Organization (NTRO), New Delhi, India for funding this work. I would like to thank Network System Laboratory, IIT Madras and Ramanujan Computing Centre, Anna University, Chennai for providing network infrastructure to test our SSE Video Conference Tool. Finally we thank our chief Co-ordinator and other all co-ordinators of this project for giving useful suggestion.

**T. Ruso** received the B.Sc degree from Department of Computer Science, Aditanar College of Arts and Science, Tiruchendur, Tamilnadu, India, in 2002, the M.Sc degree from Department of Computer Science, Annamalai University, Tamilnadu, India, in 2005, the M.E degree from Department of Computer Science and Engineering, Crescent Engineering College, Chennai, India, in 2007. He is currently pursuing Ph.D degree in Anna University, Chennai, India. His research interests include Application Layer Overlay Multicast, Multiparty Video Conference and Adhoc Network.

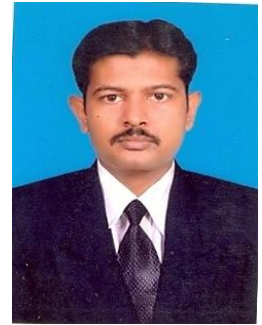

**C. Chellappan** received his Ph.D in the area of Database Systems from Anna University, Chennai, Tamil Nadu, India where he is currently the Professor of Computer Science and Engineering Department. His research interest includes Mobile Computing, Sensor Networks and Parallel system scheduling. He is also a Principal Investigator of project(Rs One Crore) on "Basic directed Collaborative Research in Smart and Secure Environment " sponsored by National Technical Research Organization, New Delhi, Govt of India, 2007-2011.

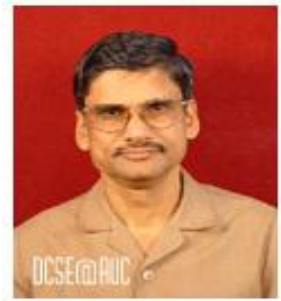